\documentclass{article}

\usepackage{arxiv}

\usepackage[utf8]{inputenc} 
\usepackage[T1]{fontenc}    
\usepackage{url}            
\usepackage{booktabs}       
\usepackage{nicefrac}       
\usepackage{microtype}      
\usepackage{lipsum}		
\usepackage{graphicx}
\usepackage{natbib}
\usepackage{doi}
\usepackage{amsmath,amsfonts,amssymb}
\usepackage{setspace}
\usepackage{tocloft}
\usepackage{float}
\usepackage[table]{xcolor}
\usepackage{macro_ref}
\usepackage{caption}
\usepackage{subcaption}
\usepackage{lineno}
\usepackage{soul}

\title{Asgard/NOTT: Cryogenic characterization of the mid-infrared chip}

\author{}

\date{}

\renewcommand{\shorttitle}{Garreau et al.: Asgard/NOTT: Cryogenic characterization of the mid-infrared chip}

\hypersetup{
pdftitle={Asgard/NOTT: Cryogenic characterization of the mid-infrared chip},
pdfsubject={astro-ph.IM},
pdfauthor={Germain Garreau},
pdfkeywords={instrumentation; integrated optics; nulling interferometry; Very Large Telescope Interferometer; Asgard/NOTT},
}

\begin{document}
\maketitle
\vspace{-6em}
\begin{center}
G. Garreau$^{a}$\footnote{ggarreau@phys.ethz.ch},
D. Defrère$^{b}$,
R. Laugier$^{b}$,
P. Chingaipe$^{b}$,
M-A. Martinod$^{c}$,
T. Mattheussen$^{b}$,
K. Missiaen$^{b}$,
J. Morren$^{b}$,
G. Raskin$^{b}$,
M. Salman$^{b}$,
W. Verstraeten$^{b}$,
A. Bigioli$^{b}$,
S. Ertel$^{d,e}$,
S. Gross$^{f}$,
X. Haubois$^{g}$,
M. Ireland$^{h}$,
A. P. Joó$^{i}$,
S. Kraus$^{j}$,
L. Labadie$^{k}$,
S. Madden$^{h}$,
F. Martinache$^{c}$,
A. Mazzoli$^{l}$,
G. Medgyesi$^{i}$,
A. Sanny$^{g}$,
N. Schuhler$^{g}$,
J. P. Scott$^{d}$,
T. A. Stuber$^{d}$, and
B. Vandenbussche$^{b}$
~\newline

$^a$ETH Zürich, Institute for Particle Physics \& Astrophysics, Wolfgang-Pauli-Str. 27, 8093 Zürich, Switzerland\\
$^b$Institute of Astronomy, KU Leuven, Celestijnenlaan 200D, 3001 Leuven, Belgium\\
$^c$Université Côte d’Azur, Observatoire de la Côte d’Azur, CNRS, Laboratoire Lagrange, France\\
$^d$Department of Astronomy and Steward Observatory, The University of Arizona, 933 North Cherry Avenue, Tucson, AZ 85721, USA\\
$^e$Large Binocular Telescope Observatory, The University of Arizona, 933 North Cherry Avenue, Tucson, AZ 85721, USA\\
$^f$MQ Photonics Research Centre, School of Mathematical and Physical Sciences, Macquarie University, NSW, 2109, Australia\\
$^g$European Southern Observatory, Alonso de Cordova 3107 Vitacura, 19001 Santiago, Chile\\
$^h$Research School of Astronomy \& Astrophysics, Australian National University, Canberra, ACT 2611, Australia\\
$^i$Konkoly Observatory, 	HUN-REN Research Centre for Astronomy and Earth Sciences, Budapest, Hungary\\
$^j$Department of Physics and Astronomy, University of Exeter, Stocker Road, Exeter EX4 4QL, UK\\
$^k$I. Physikalisches Institut, Universität zu Köln, Zülpicher Str. 77, 50937 Köln, Germany\\
$^l$STAR Institute, University of Liège, 19C allée du Six Août, 4000 Liège, Belgium
\end{center}
~\vspace{-0.5em}

\begin{abstract}
    NOTT is part of the new visitor instrument suite Asgard for the Very Large Telescope Interferometer (VLTI), and the first long-baseline nulling interferometer that will be operational in the southern hemisphere. It is an L'-band (3.5-4$\,\mu$m) instrument optimized for imaging hot exozodiacal dust and young giant planets orbiting around the snowline of nearby main-sequence stars. For planet imaging, the L’ band has the advantage of relaxing the requirements on the star-planet contrast to $\sim10^{-5}$ while limiting the level of background noise compared with longer wavelengths. Nulling interferometry in the L’-band was made possible by the development of mid-infrared integrated optics with high throughput. NOTT uses a photonic beam combiner made of Gallium Lanthanum Sulfide (GLS), manufactured at Macquarie University and characterized at ambient temperatures at Universität zu Köln. This first characterization showed that the chip could achieve the broadband contrast requirement for exoplanet imaging.
    Using the test bench of the NOTT instrument assembled at KU Leuven, and its test cryostat, we successfully cooled the chip down to $\sim$138\,K and performed its first characterization at cryogenic temperatures. The results show a raw broadband contrast of $\sim$1\,\%, similar to the previous measurements done at ambient temperatures. The splitting ratios of the different couplers are also shown to remain stable at cryogenic temperatures, with less than $\sim$2\,\% uncertainty compared to ambient measurements. These results thus show that the beam-combining properties and splitting ratios are behaving as expected at 138\,K. The current maximum throughput of the chip is estimated at $\sim$37\,\%. Future work will investigate an anti-reflection coating to reduce its Fresnel losses and increase its throughput to $\sim$50\,\%.
\end{abstract}

\keywords{instrumentation; integrated optics; nulling interferometry; Very Large Telescope Interferometer; Asgard/NOTT}

\section{INTRODUCTION}\label{sec:introduction}
    Nulling interferometry is an interferometric technique proposed in 1978 for exoplanet imaging in the infrared \citep{bracewell_detecting_1978}. Its potential is to leverage long baselines to resolve the orbit of planetary companions, while destructively interfering the stellar light to enable high-contrast imaging. Up to now, nulling interferometers -- or ``nullers" -- have only been capable of observing stellar companions and debris disks \citep{Hinz1998, Mennesson2011, Mennesson2013, Mennesson2014, Ertel_2020, martinod_scalable_2021}.
    Asgard \citep[][Martinod et al., 14148-6 in these proceedings]{Martinod2023} is a new visitor instrument for the Very Large Telescope Interferometer (VLTI), that is being assembled since 2025. NOTT is a part of this visitor instrument suite as the first long baseline nuller in the southern hemisphere, and the first nuller expected to image exoplanets, specifically young giant planets orbiting near the snowline of nearby main-sequence stars \citep{defrere_hi-5_2018, defrere2022}. The instrument will also detect and characterize hot exozodiacal dust (see Scott et al. 14148-70 in these proceedings).

    The goal of Asgard/NOTT is to achieve a planet-star contrast of $10^{-5}$ after post-processing over the 3.5-4$\,\mu$m waveband and within the nuller inner working angle \citep{Laugier2022}. The instrument will measure the spectrum of the exoplanet to characterize its upper atmosphere. To reach this level of contrast, NOTT uses a photonic chip made of Gallium Lanthanum Sulfide (GLS) glass to perform the achromatic beam combination \citep{gretzinger_towards_2019, Sanny2022}. It uses a double-Bracewell architecture that allows for internal modulation of phase and amplitude errors \citep{AngelWolf1997,Mennesson2005}. A first characterization was done at Universität zu Köln at ambient temperatures from 3.65\,$\mu$m to 3.85\,$\mu$m \citep[][Sanny et al., 14148-8 in these proceedings]{Sanny2026}. The goal of this work is to test the performance of the chip at cryogenic temperature, in terms of splitting of the light and null ratio, using the test bench developed at KU\,Leuven.

    The warm optical design of Asgard/NOTT has been implemented in a test bench at KU\,Leuven \citep{Garreau2024,Garreau2024b}, as a benchmark in preparation for NOTT's integration in Asgard in 2026. This study reports experiments performed in summer 2025, without spectral dispersion and prior to the installation of the final cryostat.  
    Latest updates on the Asgard/NOTT testbed are also presented in these proceedings, with its first dispersed null measurements at ambient temperatures (Mattheussen et al. 14148-73), and its cold nulling camera (Chingaipe et al. 14148-80). 
    Section\,\ref{sec:testbed description} describes the status of the testbed in summer 2025, and the architecture of the photonic chip. Section\,\ref{sec:cryo results} presents the results of the cryogenic measurements and the broadband characterization of the chip.

\section{Description of the setup}\label{sec:testbed description}
\subsection{Description of the KU\,Leuven testbed}\label{sec:kuleuven testbench}
\begin{figure}
    \centering
    \includegraphics[width=0.8\linewidth]{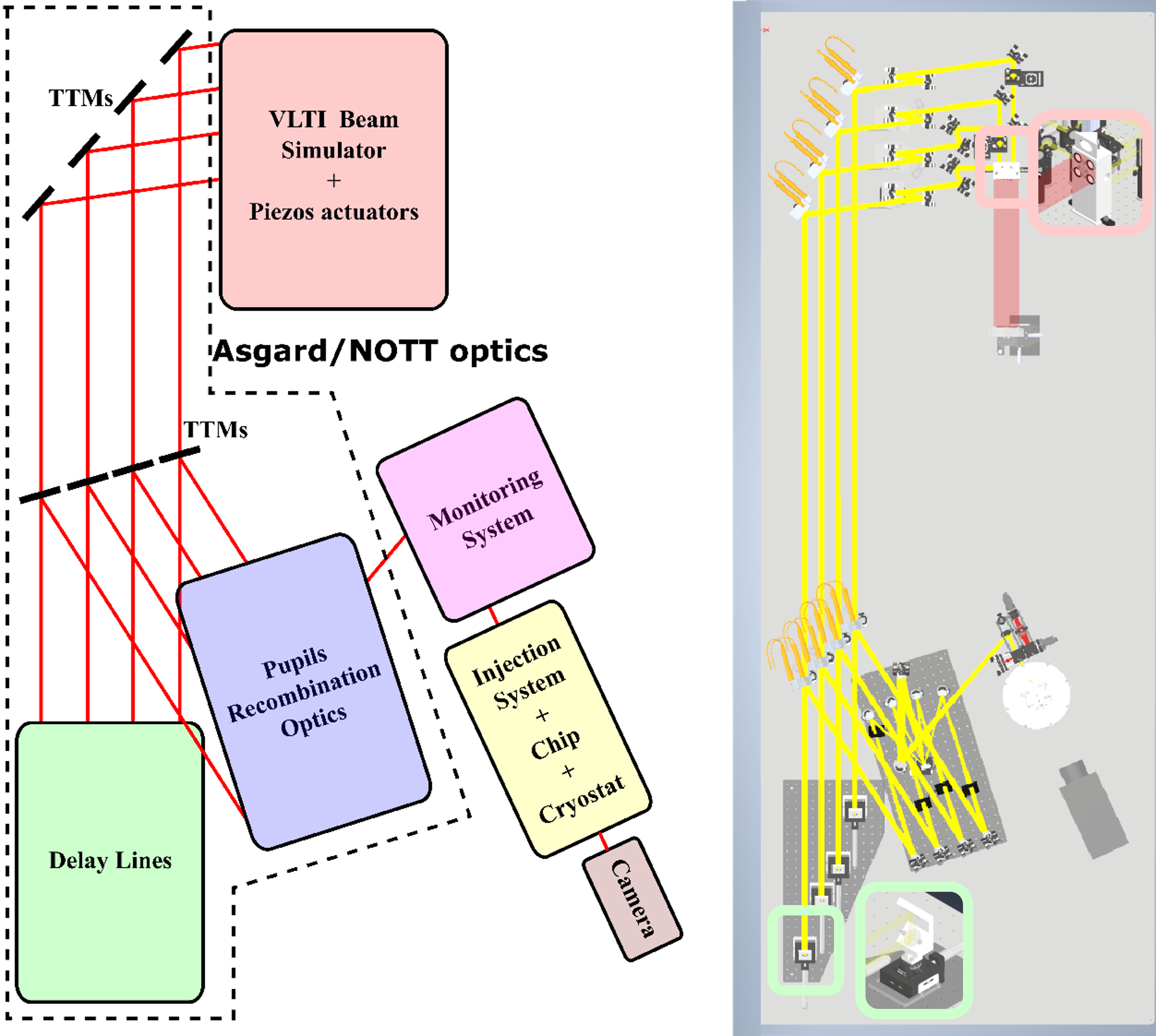}
    \caption{Left: Schematic layout of the Asgard/NOTT test bench at KU Leuven. The Asgard/NOTT warm optics (dashed lines) represents the part of the test bench that will be used identically during the Asgard assembling at Paranal. The main difference is the use of two additional mirrors in the test bench to fold the assembly and fit it on the KU Leuven optical table. The injection system is the telecentric lens used with an aperture stop to inject the light into the entrances of the chip. After propagating and interfering inside the chip, beams are emitted by the different outputs and re-imaged on the infrared camera. The monitoring system is located just before the injection system. Combined with the  tip/tilt mirrors (TTMs), it helps with the alignment of the four beams both in the image and pupil planes using the visible part of the light. Right: Top view of the opto-mechanical design of the test bench. The VLTI beam simulator (circled in pink) reproduces the four beams as expected from Heimdallr and Solarstein. Circled in green is a 3D view of a delay line. The beams are reflected back 40 mm lower toward the second TTMs. The red dotted arrow shows the expected path of the combined beams on the Paranal optical table (without the two additional mirrors). Adapted from \cite{Garreau2024b}.}
    \label{fig:NOTT_layout}
\end{figure}

\begin{figure}
    \centering
    \includegraphics[width=0.5\linewidth]{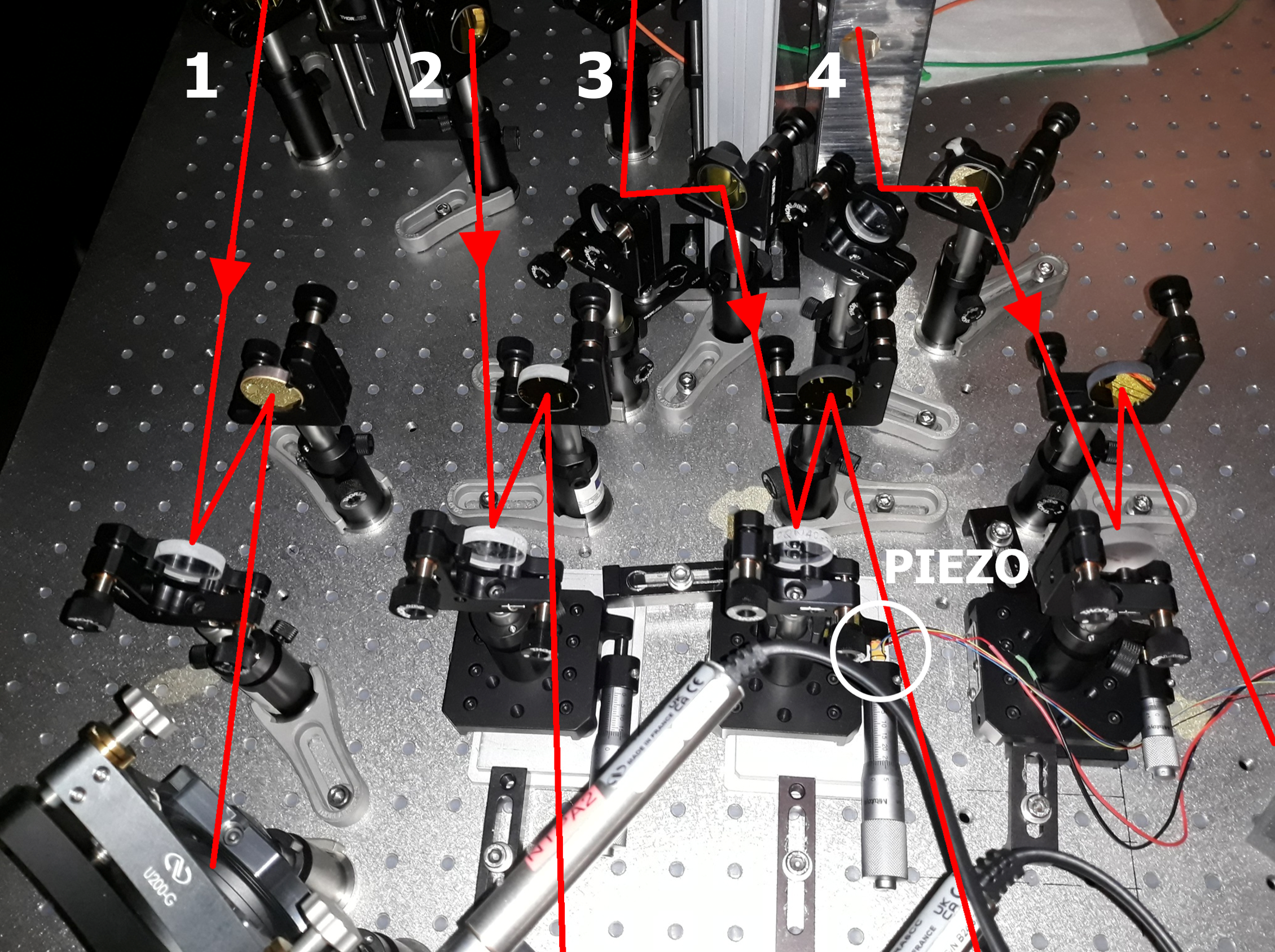}
    \caption{Image of the four additional pairs of mirrors added between the VLTI beam simulator and the first TTMs. Three of these mirrors are placed on translation stages that are compatible with piezo actuators. The numbering of the four beams is indicated on the picture in white, in red is the trajectory of the beams. One actuator is installed on the translation stage corresponding to beam 3. Taken from \cite{GarreauPhD}.}
    \label{fig:piezo}
\end{figure}

\begin{figure}
    \centering
    \includegraphics[width=0.35\linewidth]{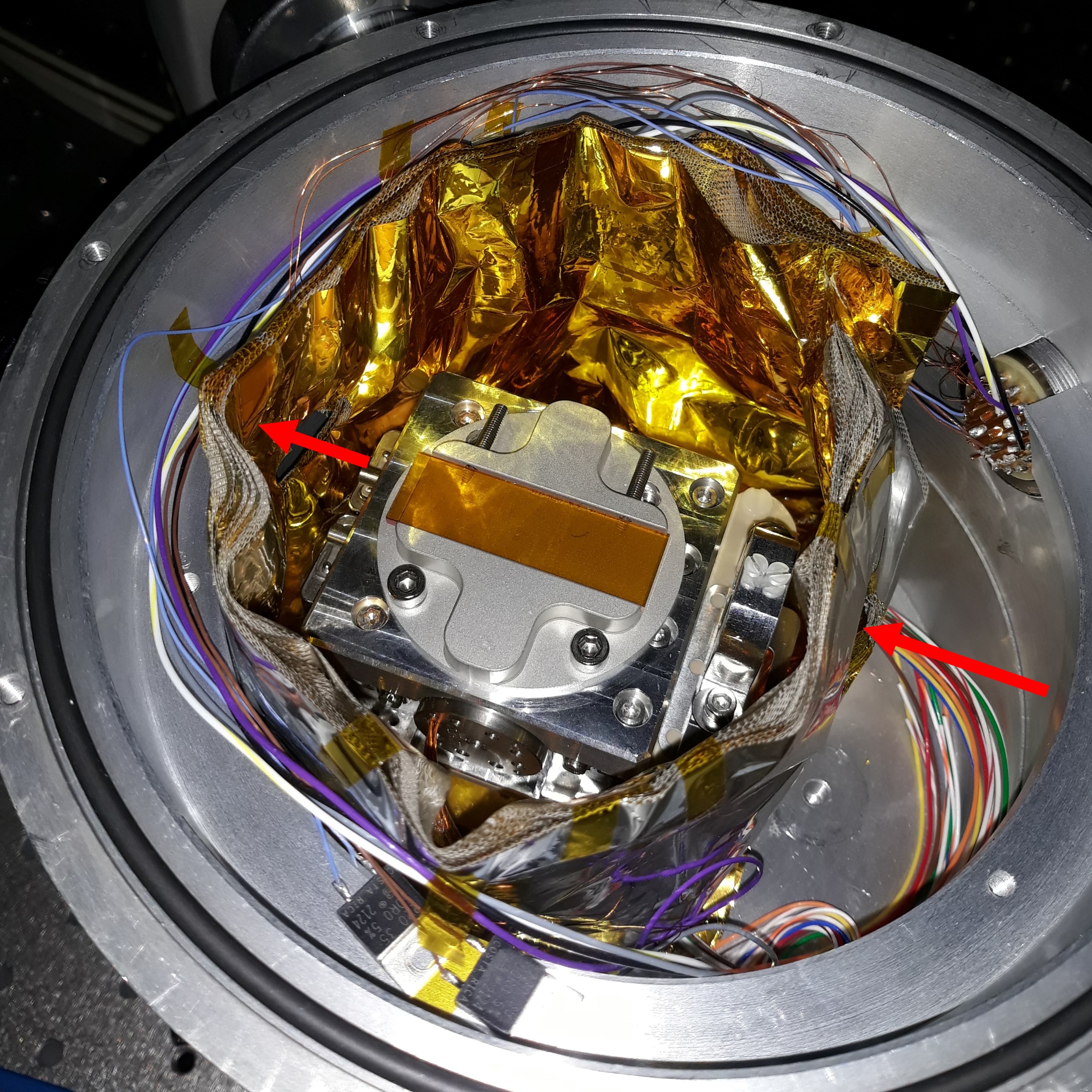}
    \includegraphics[width=0.35\linewidth]{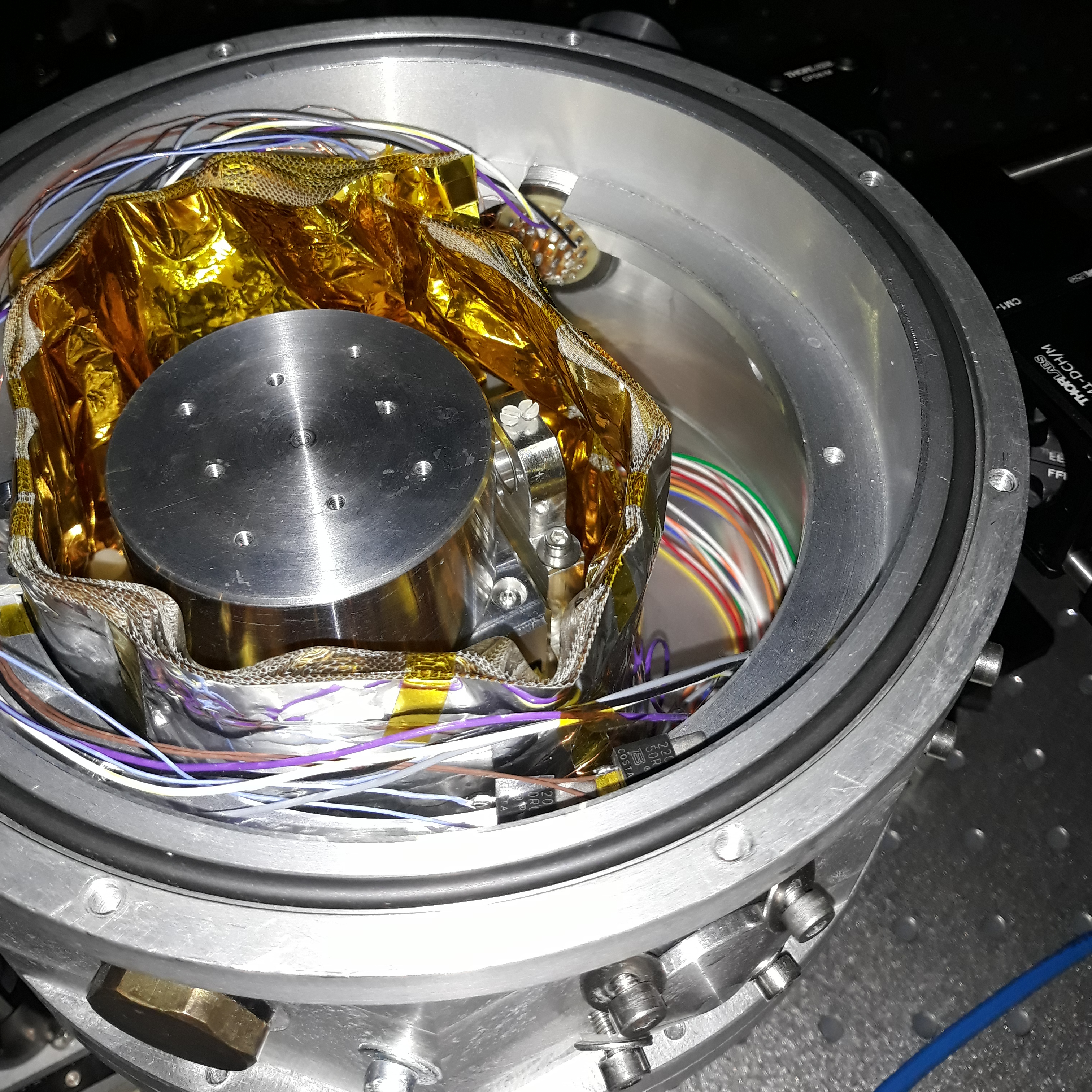}
    \includegraphics[width=0.35\linewidth]{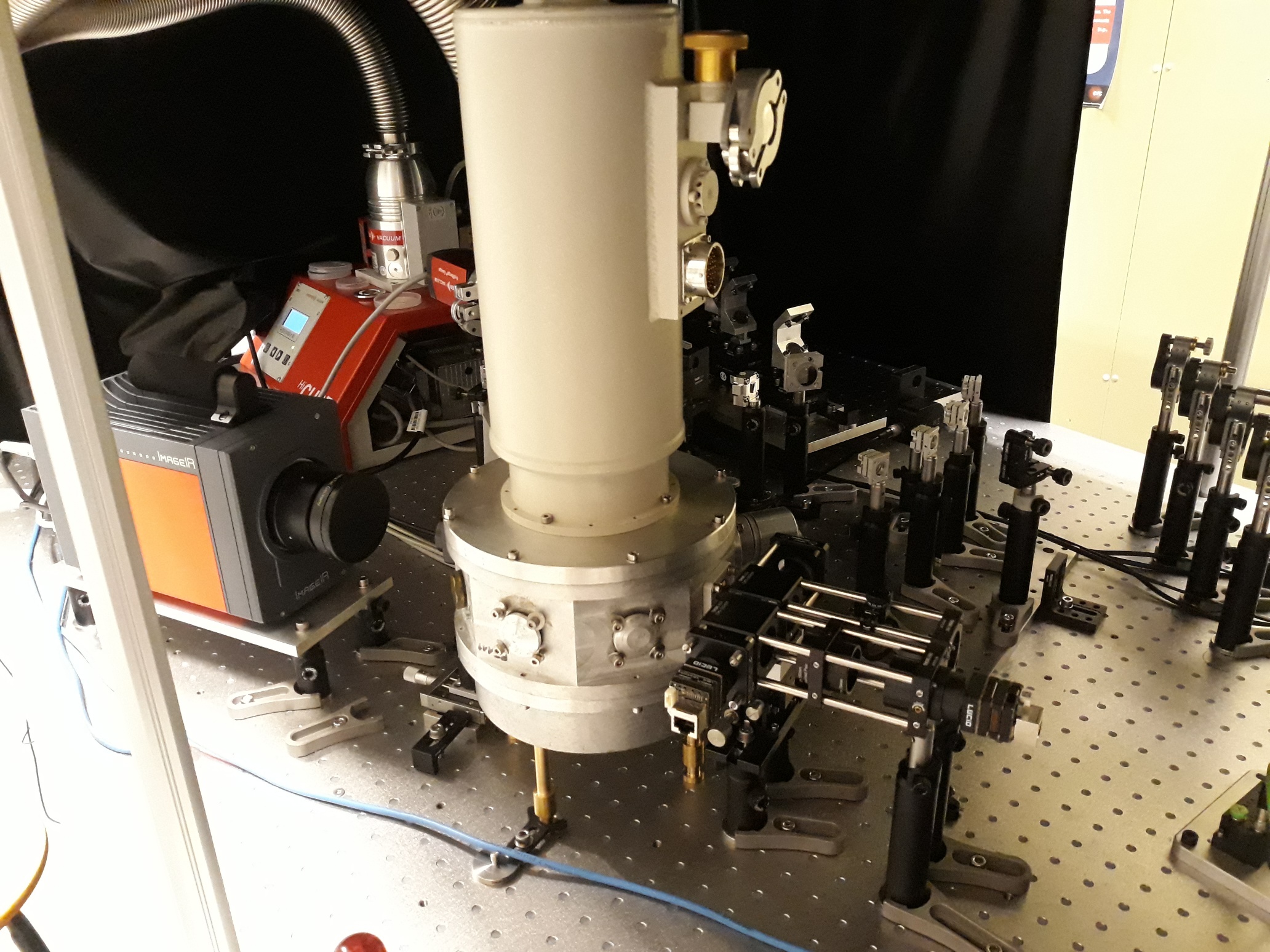}
    \includegraphics[width=0.35\linewidth]{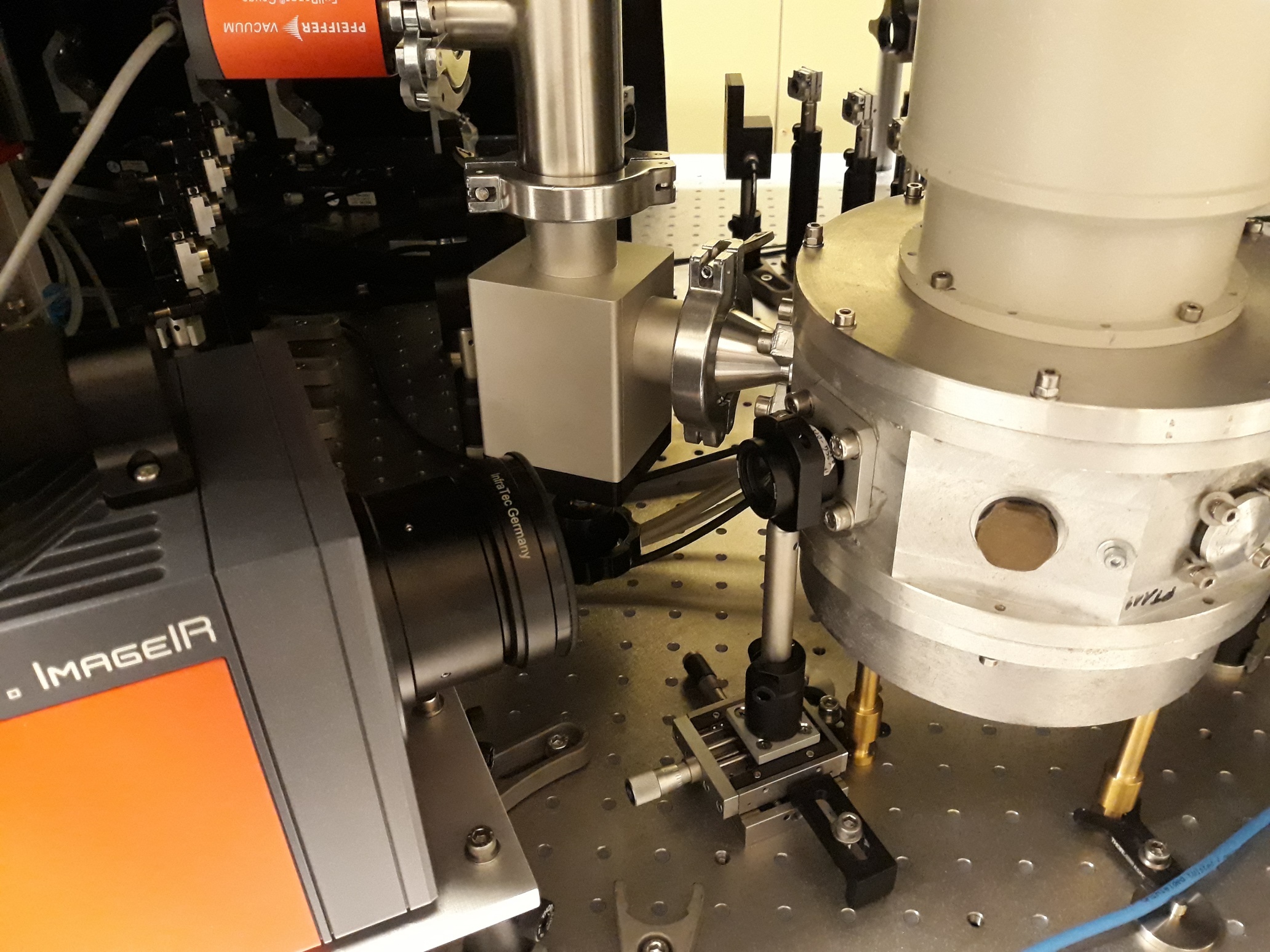}
    \caption{Top: Inside view of the cryostat with the injection lens, the cold shield, and the chip on its translation stage without (left) and with its enclosure (right). The enclosure is used to connect the cold fingers with the chip, and to fix the temperature sensors. In red is the direction of propagation of the beams. Bottom: View of the closed cryostat from its entrance with the monitoring system (left) and from its exit with the lens that collimates the outputs from the chip (right). Taken from \cite{GarreauPhD}.}
    \label{fig:test cryostat}
\end{figure}

A schematic layout of the testbed used at the time of the experiment is presented in Fig.\,\ref{fig:NOTT_layout}. Compared to the previous presentation of the test bench in \cite{Garreau2024b}, two major changes have been implemented:
\begin{itemize}
    \item \textbf{Piezoelectric ``piezo'' actuators} The strategy that was previously developed in \cite{Garreau2024b} to locate and move the delay lines to the null position could not be used anymore due to a problem of time synchronization between the infrared camera and the delay lines. The origin of this desynchronization is still unknown. To overcome this difficulty, four pairs of mirrors were added between the VLTI beam simulator and the first tip/tilt mirrors (TTMs), with three mirrors installed on a translation stage where piezos can be installed. Figure\,\ref{fig:piezo} shows the four pairs of mirrors with a piezo installed on the translation stage corresponding to beam 3. At the time of the experiment, only the piezo for beam 3 was installed and used as a fine delay line.
\\
    \item \textbf{Test cryostat} Before the delivery and installation of the final NOTT cryostat \citep{Dandumont2022spectro}, a test cryostat was made at KU\,Leuven to allow for limited cryogenic tests of the photonic chip. Its volume allows to place the chip and its mount, the injection lens, and the cold stop inside. Since the injection lens is not guaranteed at cryogenic temperatures, only the chip and its mount are cooled down. The test cryostat was implemented in 2025, along with the final Asgard/NOTT chip, and could reach a temperature down to 138\,K and $\leq$$10^{-6}$\,mbar vacuum. The top panels of Fig.\,\ref{fig:test cryostat} show the inside of the test cryostat with the chip, the injection lens, and the cold shield. The bottom panels of Fig.\,\ref{fig:test cryostat} show the closed cryostat and its installation on the optical bench.
\end{itemize}


The alignment and the injection are more challenging when the chip is placed inside a cryostat. Its position with regards to the optical axis of the injection lens is indeed critical and cannot be fine-tuned. Only the TTMs and the folding mirrors can be used to correct any misalignments, with a limited range.
If we consider a minimum of 30\,\% in coupling efficiency for tolerancing the positioning of the chip, the misalignment between the beams and the waveguide's entrances must be $<$10\,µm. When introducing correction with the TTMs, they can recover the optimal coupling efficiency up to 50\,µm misalignment.
For larger misalignments, the TTMs can be used in combination with two folding mirrors located upstream of the monitoring system to recover up to $\sim$250\,µm. This relaxes the constraints for the alignment in the test bench. However, because the folding mirrors will be removed at Paranal, the manufacturing of the final cryostat needs a precision better than 50\,µm in the positioning of the chip and the injection lens to ensure optimal coupling efficiencies as estimated in \cite{Garreau2024}. 

The focal distance between the chip and the lens is also critical. Considering again a minimum coupling efficiency of 30\,\% gives a tolerance of 300\,$\mu$m for the distance. This error cannot be corrected by the upstream optics, so it must be adjusted mechanically. The thermal expansion of the chip also needs to be accounted to ensure the optimal coupling efficiency at 138\,K for the test cryostat.

In the test bench, injection could be obtained at ambient temperature using the TTMs with the folding mirror, and coarsely adjusting the focal distance. Using a narrow L'-band filter (3.62-3.98\,$\mu$m), an output intensity up to 150\,Digital Values (DV) was obtained on the camera for the brightest outputs. When comparing with the previous setup in \cite{Garreau2024b}, the brightest outputs were up to $\sim$250\,DV. 
Using the previous throughput estimate that was calculating the coupling efficiency of the beams into the chip at $\geq$60.4\,\% \cite{GarreauPhD}, this then corresponds to a drop of coupling efficiency to $\geq$36\,\%, assuming the losses in throughput mostly come from it.
The positioning errors between the chip and the injection lens are likely the origin of this drop. To counteract it and improve the S/N of the outputs, a wider L'-band filter (3.4-4.0\,µm) is used for the characterization of the chip and the null measurements.

\subsection{Chip architecture}
\begin{figure}
    \centering
    \includegraphics[width=0.75\linewidth]{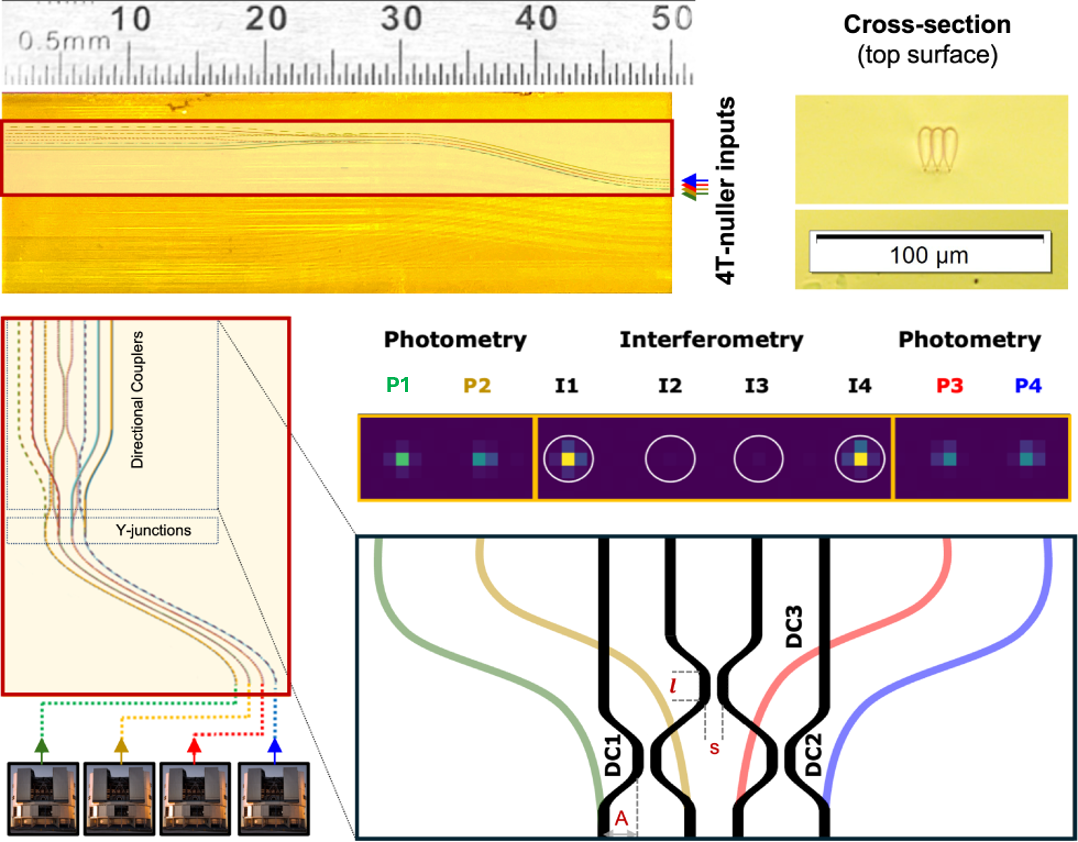}
    \caption{Layout of the NOTT chip. Top left: top view of the NOTT chip with visible waveguides and arrows indicating the positions of the four inputs. Top right: cross section of a triplet waveguide. Bottom left: schematic layout of a nuller with its four inputs, its side-steps, and its three directional couplers combining the light. Bottom right: schematic layout of the eight outputs of a nuller, with the four photometric outputs $P_{1-4}$ and the four interferometric outputs $I_{1-4}$. In the double-Bracewell mode, $I_1$ and $I_4$ are the constructive outputs, while $I_2$ and $I_3$ are the nulled signal. Adapted from \cite{Sanny2026}.}
    \label{fig:Sanny fig1}
\end{figure}

The NOTT chip used in this experiment is the third generation of GLS chip manufactured at Macquarie University, and the one that will be integrated at the VLTI. Its architecture is described in Fig.\,\ref{fig:Sanny fig1}.

\begin{figure}
    \centering
    \includegraphics[width=0.5\linewidth]{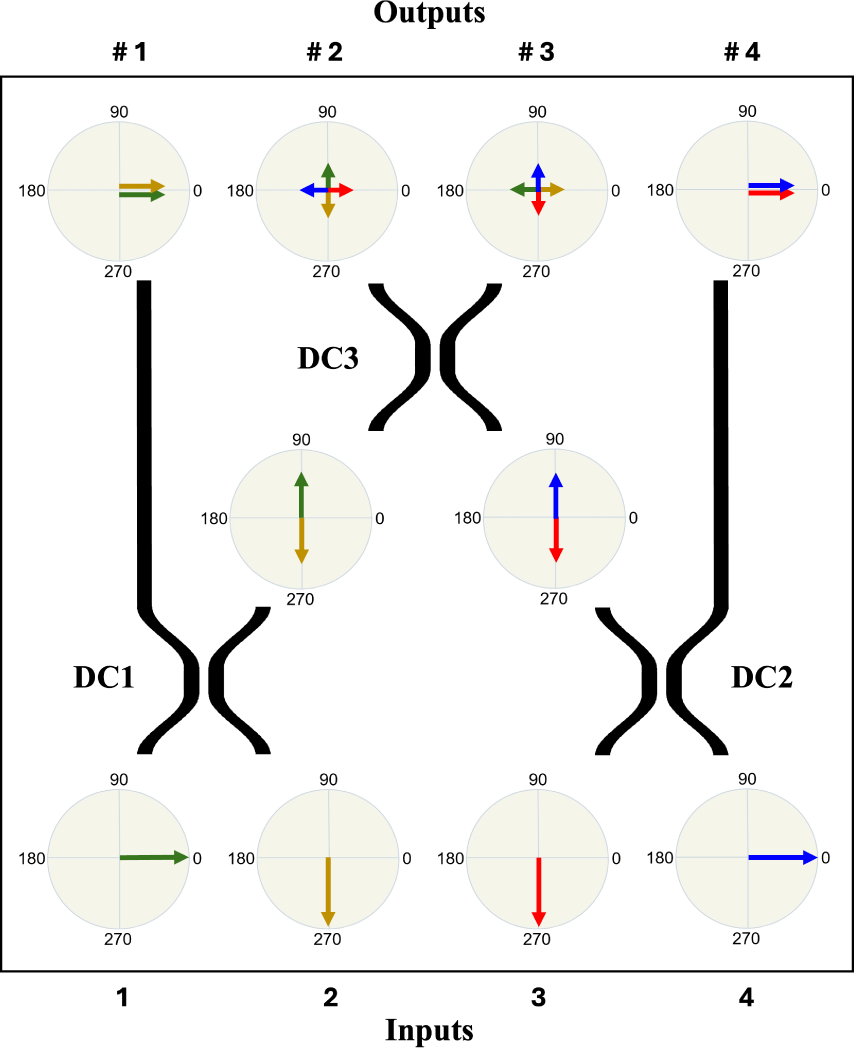}
    \caption{Functional phasor diagram of the double-Bracewell nuller with complex amplitudes of inputs 1 (green), 2 (yellow), 3 (red), and 4 (blue). DC1, DC2, and DC3 are three directional couplers with a 50/50 splitting ratio in intensity. DC1 and DC2 produce constructive outputs at \#1 and \#4, while the two nulled outputs are feeding DC3. The nulled outputs \#2 and \#3 feature two transmission maps, which are centro-symmetric to one another with regard to the central star. Taken from \cite{Sanny2026}.}
    \label{fig:sanny fig2}
\end{figure}

The final Asgard/NOTT chip was delivered at the beginning of 2025 by the Universität zu Köln. The chip has five nulling blocks following the same double-Bracewell architecture as shown in Fig.\,\ref{fig:Sanny fig1}. The interaction length of the directional couplers (DCs) is going from 6\,mm to 8\,mm by steps of 0.5\,mm. The first characterization of the chip at warm temperature \citep[][Sanny et al., 14148-8 in these proceedings]{Sanny2026} showed that the most optimal nuller corresponds to an interaction length of 7.5\,mm, with a 60/40 splitting ratio for DC1 and DC2, and a 50/50 splitting ratio for DC3 (see Fig.\,\ref{fig:sanny fig2} for a detailed view of the functional phasor diagram for the double-Bracewell mode of the nuller).

\section{Results}\label{sec:cryo results}

After optimizing the injection with the chip placed inside the test cryostat, a broadband (3.4-4.0\,$\mu$m) characterization could be performed at ambient and cryo-temperatures.
The results of the chip characterization on the KU\,Leuven testbed are taken from \cite{GarreauPhD}. In this section, we present the comparison of the broadband $\kappa$-matrices of the chip and its null measurements at both ambient and cryogenic temperatures.

\subsection{$\kappa$-matrix measurements}

A first test consists of checking the splitting ratios by measuring the $\kappa$-matrix of the nuller \citep{v_coude_du_foresto_deriving_1997}. $\kappa$ is the matrix that relates the intensity measured at each output ($P_{1-4}, I_{1-4}$) with the intensity injected in the inputs (i$_1$, ..., i$_4$). This matrix is measured by injecting each of the four beams one-by-one and observing the intensity it generates in the outputs. This measure requires to correct for the difference in background emission for the different outputs, and the blackbody emission generated by the closed shutters that couples with the waveguides. 
Equation\,(\ref{eq:kappa 7.5}) shows the result of the measurement at ambient temperature. All the factors have a typical uncertainty of $<$0.02.
\begin{equation}\label{eq:kappa 7.5}
	\begin{pmatrix}
	\text{P}_1 \\ \text{P}_2 \\ \text{I}_1 \\ \text{I}_2 \\ \text{I}_3 \\ \text{I}_4 \\ \text{P}_3 \\ \text{P}_4
	\end{pmatrix}
	= \kappa_{\text{warm}} \times
	\begin{pmatrix}
	\text{i}_1 \\ \text{i}_2 \\ \text{i}_3 \\ \text{i}_4
	\end{pmatrix} =
	\begin{pmatrix}
	0.45 	& 0 	  	& 0 		& 0 \\
	0 	& 0.44 	& 0 		& 0 \\
	0.35	& 0.19 	& 0 		& 0 \\
	0.12 	& 0.22  	& 0.16 	& 0.09 \\
	0.08 	& 0.15 	& 0.21 	& 0.11 \\
	0 	& 0   	  	& 0.20 	& 0.36 \\
	0 	& 0      	& 0.43 	& 0 \\
	0 	& 0      	& 0 		& 0.44 \\
	\end{pmatrix}
	\times
	\begin{pmatrix}
	\text{i}_1 \\ \text{i}_2 \\ \text{i}_3 \\ \text{i}_4
	\end{pmatrix}
\end{equation}

The $\kappa$-matrix can also be rewritten as $\kappa'$ to show the relationship between the interferometric outputs (I$_1$,..,I$_4$) and the photometric outputs (P$_1$,..,P$_4$). Equation\,\ref{eq:kappa 7.5 PtoI} shows the rewritten matrix.

\begin{equation}\label{eq:kappa 7.5 PtoI}
	\begin{pmatrix}
		\text{I}_1 \\ \text{I}_2 \\ \text{I}_3 \\ \text{I}_4
	\end{pmatrix}
	= \kappa_{\text{warm}}' \times 
    \begin{pmatrix}
	\text{P}_1 \\ \text{P}_2 \\ \text{P}_3 \\ \text{P}_4
	\end{pmatrix}
     =
	\begin{pmatrix}
		0.77 & 0.43 & 0 & 0\\
		0.27 & 0.50 & 0.37 & 0.20 \\
		0.20 & 0.34 & 0.49 & 0.25\\
		0 & 0 & 0.47 & 0.82
	\end{pmatrix}
	\times 
	\begin{pmatrix}
	\text{P}_1 \\ \text{P}_2 \\ \text{P}_3 \\ \text{P}_4
	\end{pmatrix}
\end{equation}

Using Eq.\,(\ref{eq:kappa 7.5}), we find the splitting ratios of each beam at the DCs which are summarized in Table\,\ref{tab:splitting ratio 7.5mm}. The results from \cite{Sanny2026} show that the requirements for the different DCs should be respected with less than 5\,\% dispersion in the L'-band. This is not the case for the DC3. 
One hypothesis to explain this discrepancy is that the passband used in the previous experiment went from 3.65 to 3.85\,µm, while we use a larger L'-band filter going from 3.4 to 4\,µm. The larger bandwidth is likely to generate more chromatic dispersion of the splitting ratios. A spectral dispersion of the outputs will be necessary to confirm and characterize it.

The $\kappa$-matrices of the nuller at cryogenic temperature are given by Eq.\,(\ref{eq:kappa 7.5 cryo}). Finally, the splitting ratios of the DCs are listed in Table\,\ref{tab:splitting ratio 7.5mm}.

\begin{equation}\label{eq:kappa 7.5 cryo}
    \kappa_{\text{cryo}}
    =
	\begin{pmatrix}
	0.45 	& 0 	  	& 0 		& 0 \\
	0 	& 0.45 	& 0 		& 0 \\
	0.35	& 0.19 	& 0 		& 0 \\
	0.12 	& 0.21  	& 0.15 	& 0.08 \\
	0.08 	& 0.15 	& 0.20 	& 0.10 \\
	0 	& 0   	  	& 0.20 	& 0.37 \\
	0 	& 0      	& 0.43 	& 0 \\
	0 	& 0      	& 0 		& 0.45 \\
	\end{pmatrix}, \quad
    \kappa_{\text{cryo}}'
    =
	\begin{pmatrix}
		0.78 & 0.42 & 0 & 0\\
		0.27 & 0.47 & 0.35 & 0.18 \\
		0.18 & 0.33 & 0.51 & 0.22\\
		0 & 0 & 0.47 & 0.82
	\end{pmatrix}
\end{equation}

The $\kappa$-matrices are found to be in good agreement with the ones at ambient temperature, with $<$5\,\% difference between the individual factors. The splitting ratios are also found to be $<$5\,\% different than the ones measured at ambient temperature.

\begin{table}
    \centering
    \caption{Splitting ratios of the beams at ambient temperature (left) and cryogenic temperature (right) for the nuller of the Asgard/NOTT chip with an interaction length of 7.5\,mm.}
    \begin{tabular}{c c c}
        \hline\hline
        \textbf{Warm} & DC1/DC2 & DC3 \\ \hline
         From \cite{Sanny2026} & 60/40 & 50/50 \\ \hline
        i$_1$ & 64/36 & 60/40 \\
        i$_2$ & 66/34 & 60/40 \\
        i$_3$ & 65/35 & 57/43 \\
        i$_4$ & 64/36 & 55/45 \\ \hline
    \end{tabular}
    \quad
    \begin{tabular}{c c c}
        \hline\hline
        \textbf{Cryo} & DC1/DC2 & DC3 \\ \hline
         From \cite{Sanny2026} & 60/40 & 50/50 \\ \hline
        i$_1$ & 64/36 & 60/40 \\
        i$_2$ & 65/35 & 58/42 \\
        i$_3$ & 65/35 & 59/41 \\
        i$_4$ & 67/33 & 56/44 \\ \hline
    \end{tabular}\\
    \raggedright
    \textbf{Notes.} The left number of the splitting ratio is the portion of signal that remains in its original waveguide, while the right number is the portion that couples to the other waveguide. The results can be compared to those from the previous characterization presented in \cite{Sanny2026}.
    \label{tab:splitting ratio 7.5mm}
\end{table}

\subsection{Null measurement}
\begin{figure}
    \centering
    \includegraphics[width = \linewidth]{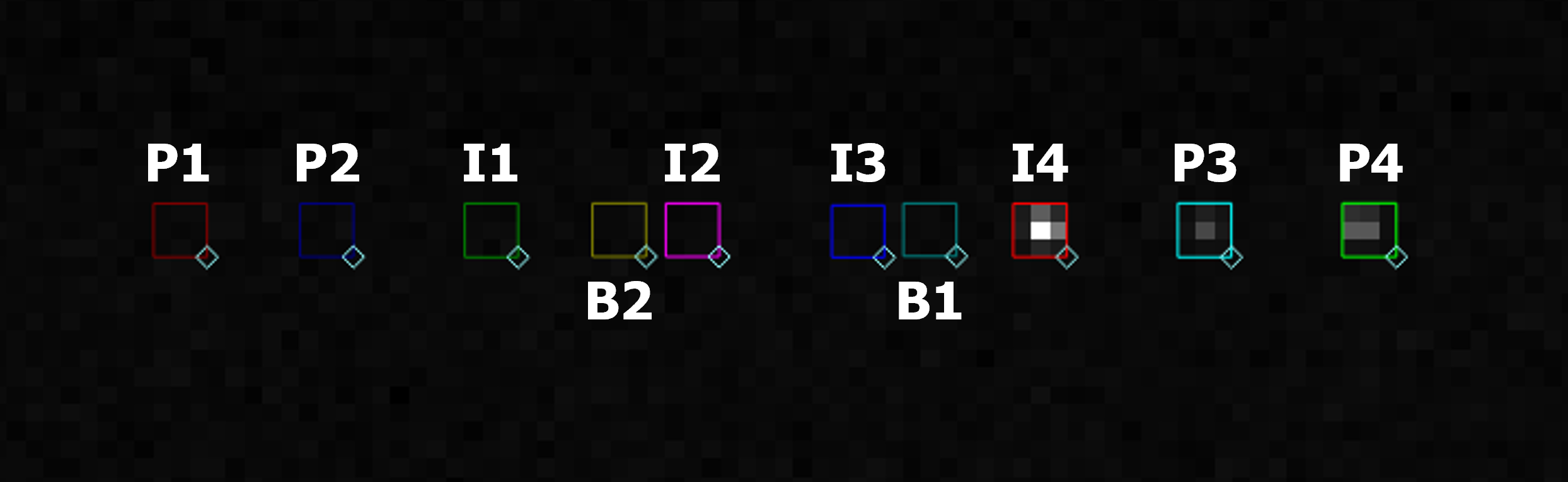}
    \caption{Image of the infrared camera with the outputs I2 and I3 at the null position. Beams 1 and 2 are switched off. B1 and B2 are the pixel regions used for background measurements.}
    \label{fig:null_image}
\end{figure}

We choose to generate the null with a two-beams recombination between beams 3 and 4. Figure\,\ref{fig:null_image} shows the image of the infrared camera with beam 1 and beam 2 switched off, the outputs I2 and I3 at the null position, and I4 at the bright position.

\begin{figure}
    \centering
    \includegraphics[width = 0.49\linewidth]{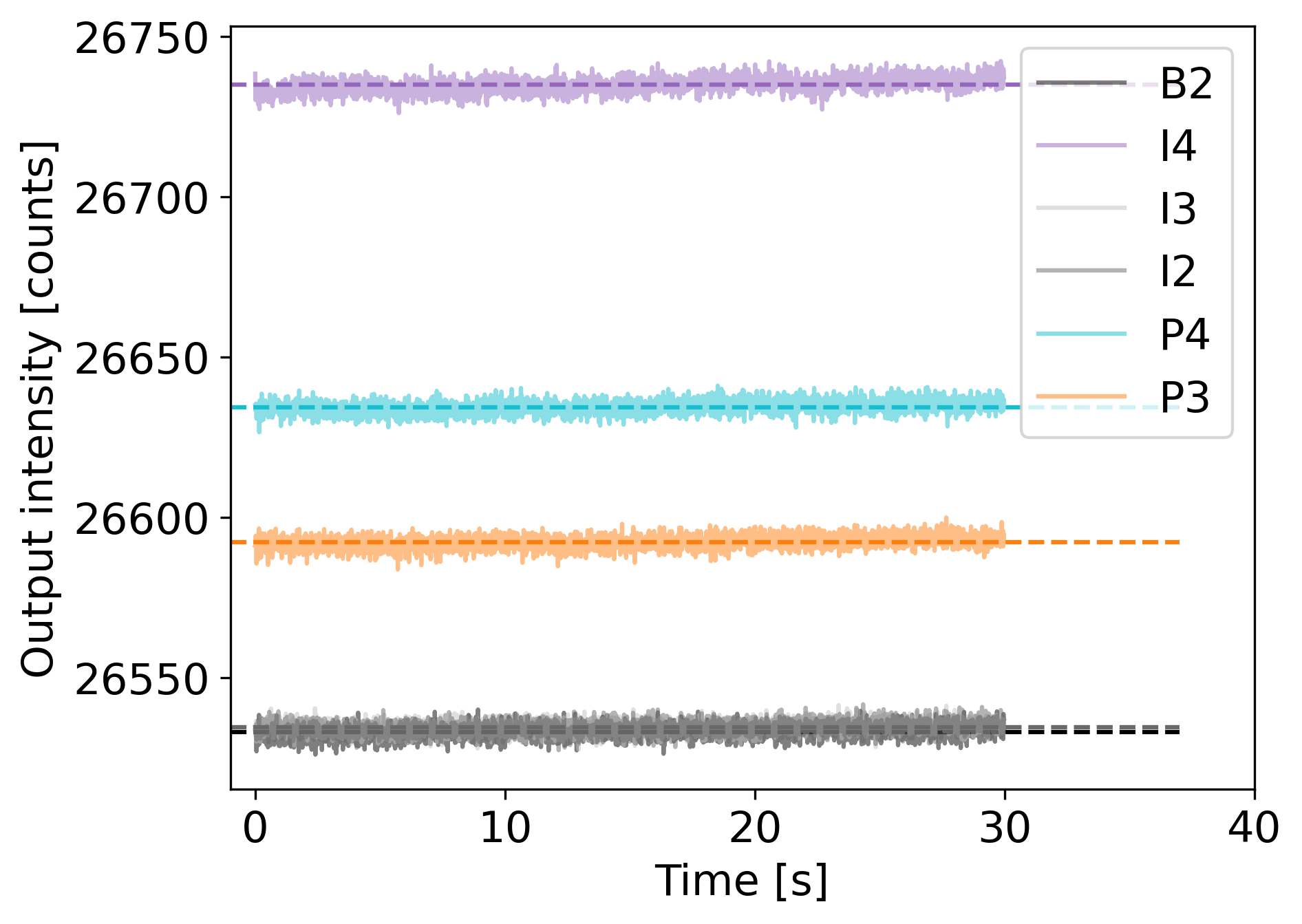}
    \includegraphics[width = 0.49\linewidth]{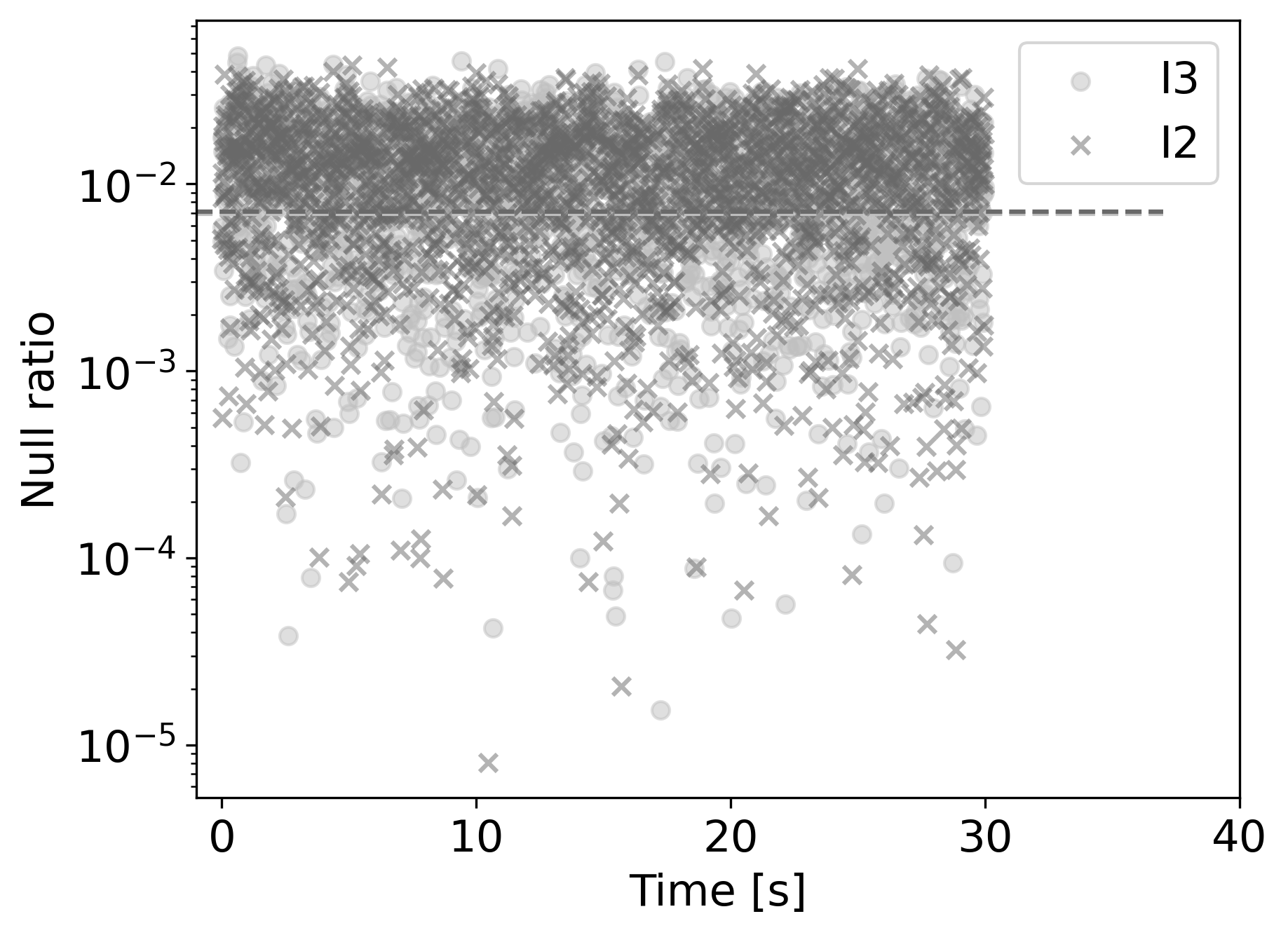}
    \caption{Results with the chip at ambient temperature. Left: Intensities of the outputs P3, P4, I2, I3, I4, and B2 recorded by the infrared camera, after correction of the differential background emission between the outputs and B2. The framerate is 100\,Hz with 3\,ms integration time. Right: Null ratios $N_2$ and $N_3$ of the outputs I2 and I3, respectively.}
    \label{fig:null_measure}
\end{figure}
\begin{figure}
    \centering
    \includegraphics[width = 0.49\linewidth]{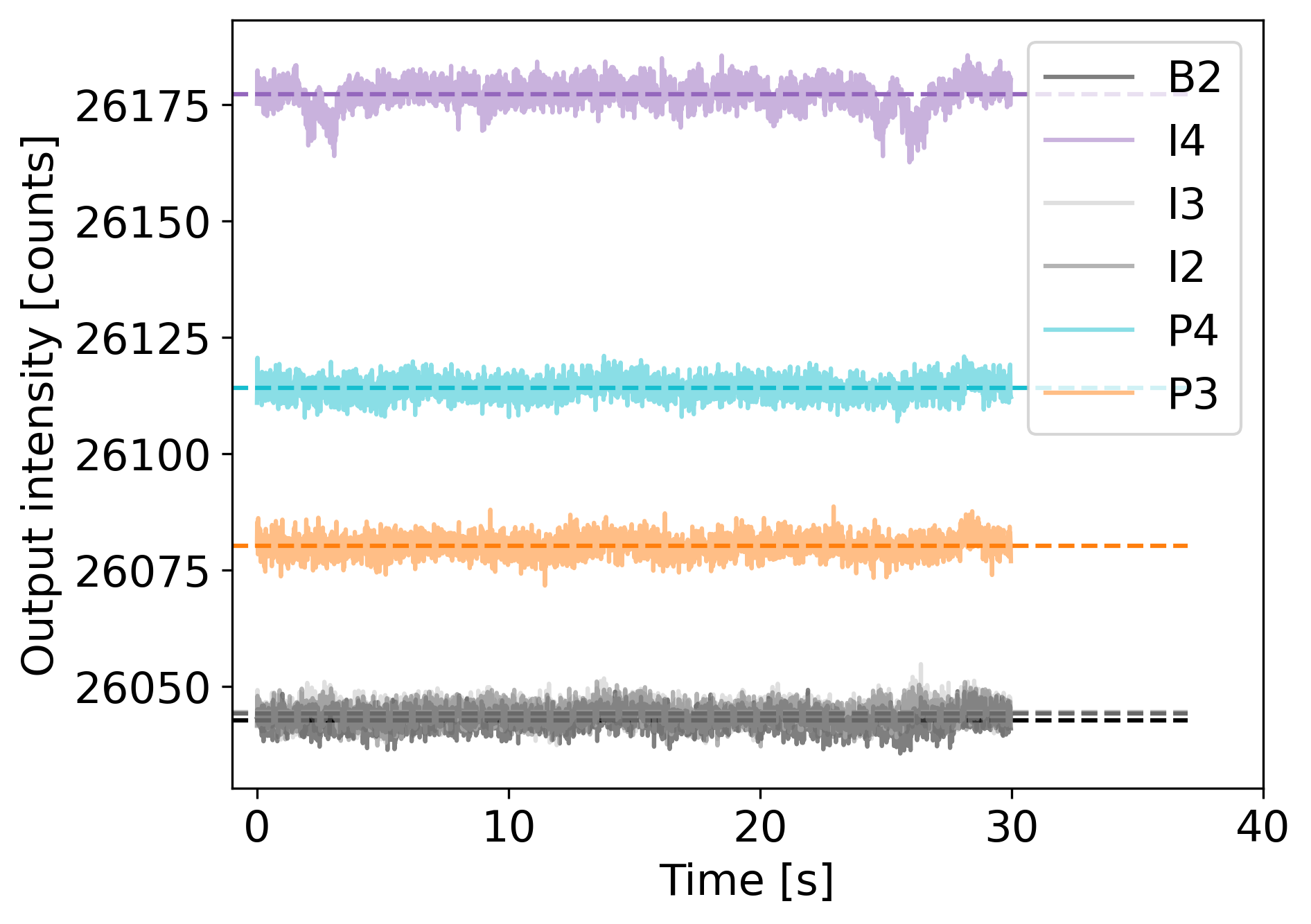}
    \includegraphics[width = 0.49\linewidth]{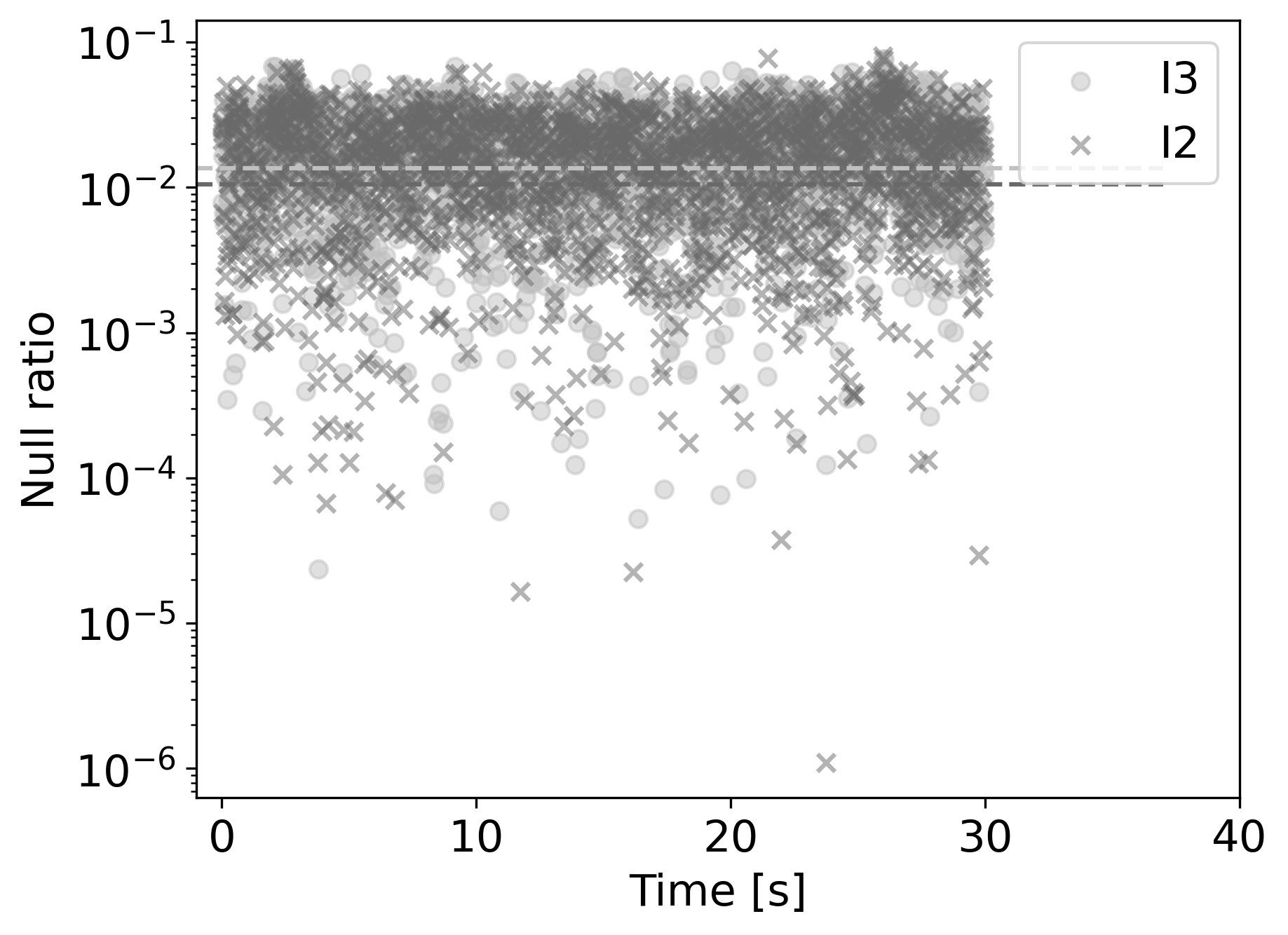}
    \caption{Results with the chip at cryogenic temperature. Left: Intensities of the outputs P3, P4, I2, I3, I4, and B2 recorded by the infrared camera, after correction of the differential background emission between the outputs and B2 . The framerate is 100\,Hz with 3\,ms integration time. Right: Null ratios $N_2$ and $N_3$ of the outputs I2 and I3, respectively.}
    \label{fig:null_measure_cryo}
\end{figure}

The outputs P3, P4, I2, I3, I4, and the background B2 are recorded during 30\,s with I2 and I3 at the null position, and with 3000 frames of 3\,ms integration time.
The recorded intensities of each output at ambient and cryogenic temperatures are shown on the left panels of Figs.\,\ref{fig:null_measure}-\ref{fig:null_measure_cryo}, respectively, after correction of the differential background emission with B2. 
We calculate the null ratios $N_2$ and $N_3$ of I2 and I3, respectively, with Eq.\,(\ref{eq:ch2 null ratios}). 
\begin{equation}\label{eq:ch2 null ratios}
N_2 = \frac{I2-B2}{I4-B2},\qquad
N_3 = \frac{I3-B2}{I4-B2}.
\end{equation}
The right panels of Figs.\,\ref{fig:null_measure}-\ref{fig:null_measure_cryo} show the null ratios $N_2$ and $N_3$ during the measurement sequence. The final null ratio $N$ considered is the sum of $N_2$ and $N_3$:
\begin{equation}
    N = N_2+N_3.
\end{equation}

\begin{figure}
    \centering
    \includegraphics[width = 0.55\linewidth]{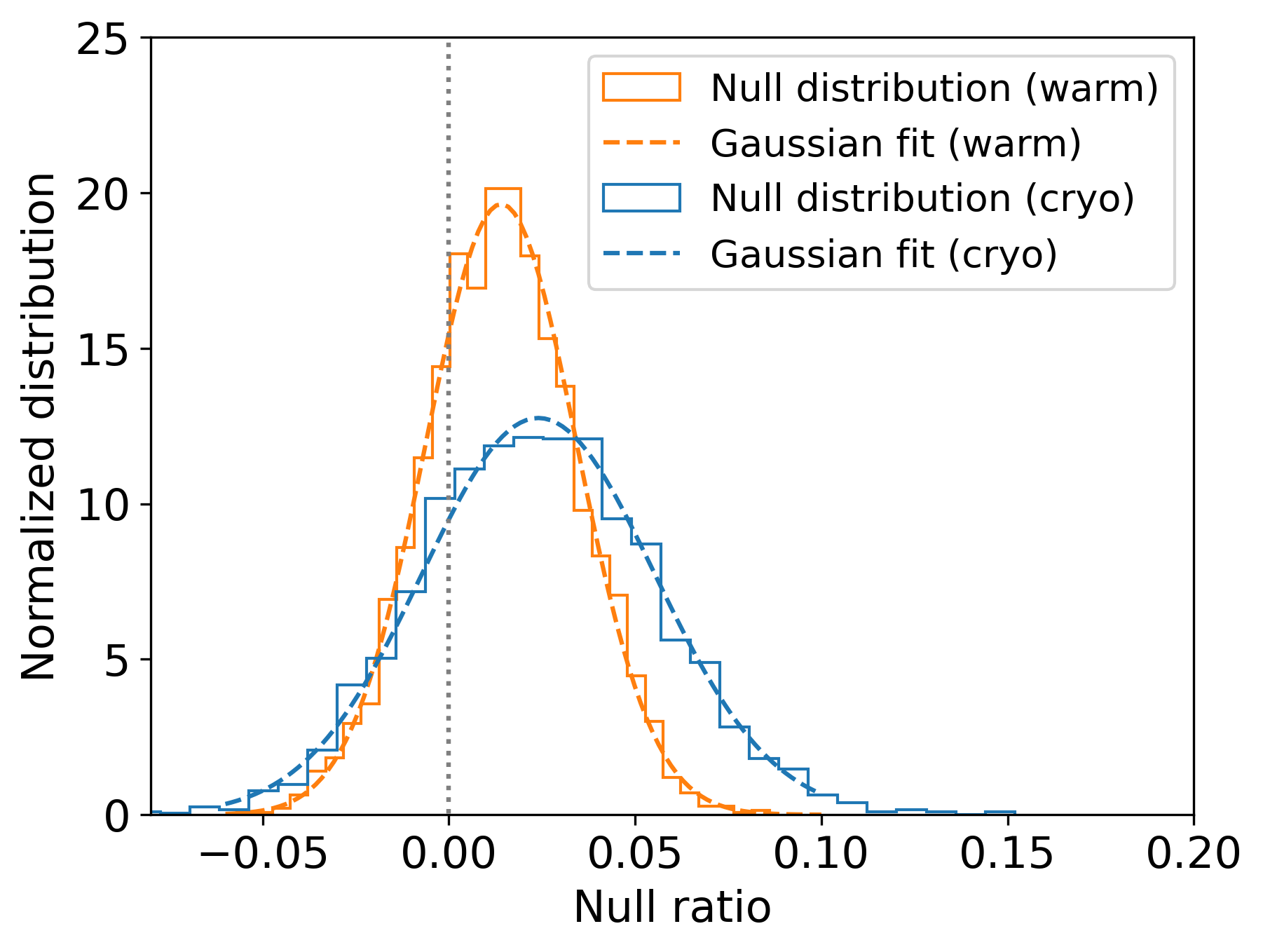}
    \caption{Normalized distribution of the null ratio $N$ with the chip at ambient (orange) and cryogenic (blue) temperatures. The continuous lines are the histogram of $N$, and the dashed lines give the best fits with a gaussian distribution. The null distribution at cryogenic temperature is broader because of larger vibration noises coming from the nearby vacuum pump.}
    \label{fig:null_hist_cryo}
\end{figure}

\paragraph{Analysis at ambient temperature} The experiment gives $N=1.4\pm2$\,\%, which is consistent with the results previously found in \cite{Sanny2026}. Figure\,\ref{fig:null_hist_cryo} shows the normalized distribution of $N$ and its best fit with a gaussian distribution. We perform a Kolmogorov–Smirnov test between the real null distribution and the best fit gaussian distribution. The null hypothesis that the two distributions are not identical is rejected with more than 95\,\% certainty, which confirms that the null distribution can be approximated to a gaussian distribution. Following this conclusion, the null stability $N_s$ during the 30\,s experiment can be approximated to:
\begin{equation}\label{eq:ch2 null stab}
    N_s = \frac{\sigma_N}{\sqrt{n}},
\end{equation}
with $\sigma_N$ the standard deviation of $N$, and n the number of data points during the sequence of measurement. This gives a null stability $N_s=3.7\times10^{-4}$ for 30\,s of measurements.

Using the mean values of P3 and P4, and Eq.\,(\ref{eq:kappa 7.5 PtoI}), the intensity mismatch between the two beams at the null outputs can be estimated. We find an output intensity of 59 and 101 after background correction for P3 and P4, respectively, which corresponds to a relative intensity error $\sigma_I$ of 3.9\,\% and 8.8\,\% for I2 and I3, respectively. The null ratio contribution from intensity mismatch $N_{\sigma_I}$ is given by
\begin{equation}
    N_{\sigma_I} = \frac{\sigma_I}{16}.
\end{equation}
For the measurement at ambient temperature, this gives $<5\times10^{-4}$, which shows that $N$ is not dominated by the contribution from intensity mismatch. 

Since the null measurement is performed with broadband light, we also take into account the loss of coherence due to the bandwidth smearing. The DCs of the chip are generating an achromatic $\frac{\pi}{2}$ phase shift between the pair of beams. This means that an additional $\frac{\pi}{2}$ is needed to reach a $\pi$ phase shift and obtain a destructive interference. This additional $\frac{\pi}{2}$ is provided by the movement of the piezo actuator which changes the OPD, and generates a chromatic phase difference. It induces a loss of coherence that depends on the bandwidth of the filter. We calculate the null ratio from this loss of coherence using the wide L'-band filter ($\lambda_0=3.7$\,µm and $\Delta\lambda=0.6$\,µm). The result is a null ratio of $1.35\times10^{-3}$. Our measured $N$ is therefore not dominated by the bandwidth smearing of the filter.

These results are summarized in Table\,\ref{tab:summary}.

\paragraph{Analysis at cryogenic temperature}
The experiment gives $N=2.4\pm3$\,\%. Fig.\,\ref{fig:null_hist_cryo} gives the normalized distribution of $N$, which we again approximate to a gaussian distribution. This gives a null stability of $N_s=5.7\times 10^{-4}$ for 30\,s of measurements. P3 and P4 have output intensities of 37 and 71, respectively, after background correction which correspond to a relative intensity error $\sigma_I$ of 0.7\,\% and 9.4\,\% for I2 and I3, respectively. The null ratio contribution of intensity mismatch is thus again $\leq5\times10^{-4}$, which is negligible compared to $N$. These results are summarized in Table\,\ref{tab:summary}.

\subsection{Estimation of coupling efficiency}
\begin{figure}
    \centering
    \includegraphics[width = 0.55\linewidth]{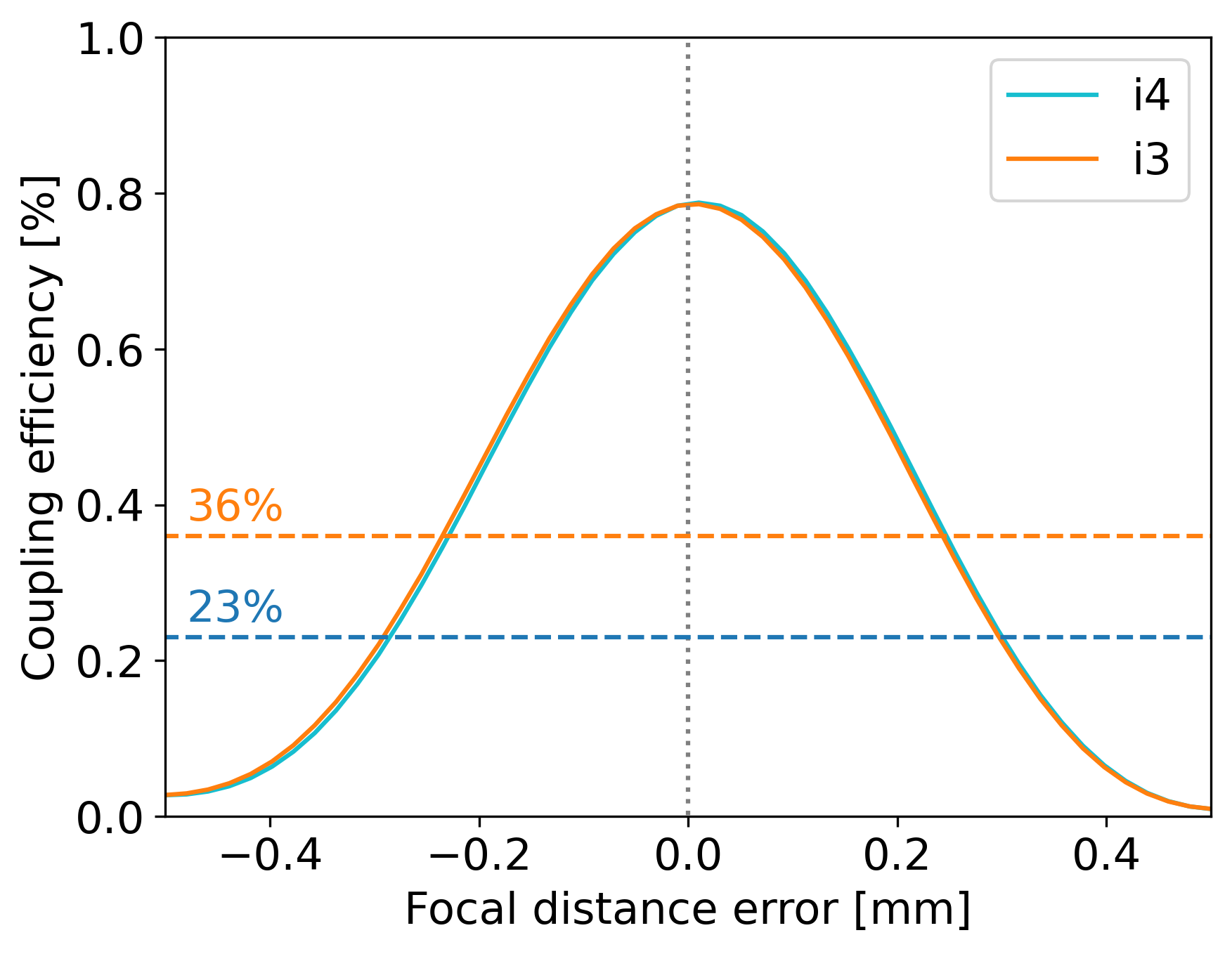}
    \caption{Simulated coupling efficiencies of the beams i3 (orange) and i4 (cyan) with regards to the error on the focal distance between the injection lens and the front face of the chip. The dashed lines indicate the estimated coupling efficiencies at ambient (orange) and cryogenic (blue) temperatures. The simulation was made using Zemax.}
    \label{fig:focal_dist_err}
\end{figure}

Using P3 and P4 and the $\kappa$-matrices obtained at ambient and cryogenic temperatures, we can also estimate the intensity of injected light in both cases and the relative loss in coupling efficiency when going to cryogenic temperatures. We find a relative loss of 63\,\% and 69\,\% for P3 and P4, respectively. Assuming the coupling efficiency at ambient temperature was 36\,\% from Sect.\,\ref{sec:kuleuven testbench}, this brings the coupling efficiency down to $\sim$23\,\% at 138\,K, which is below the arbitrary tolerancing value of 30\,\% that we chose. Figure\,\ref{fig:focal_dist_err} represents the simulated coupling efficiencies of the two input beams i3 and i4 with regards to the error on the focal distance between the injection lens and the front face of the chip. A drop of coupling efficiency from 36\,\% to 23\,\% would correspond to an increase of the focal distance error of $\sim$50\,µm. GLS has a thermal expansion coefficient of $10^{-5}$\,/K at 273\,K. Assuming this coefficient is constant at cryogenic temperature, cooling the chip with a length of 50\,mm from 300\,K to 138\,K will reduce its length by $\sim$80\,µm. If this contraction is isotropic, then the distance between the chip and the lens will increase by $\sim$40\,µm, which could explain the observed drop in coupling efficiency.
More efforts are thus required to optimize the focal distance between the injection lens and the chip, and take into account its thermal expansion.

\begin{table}
    \centering
    \caption{Summary of the null ratio measurements at warm and ambient temperatures, their null stability, and the estimation of the contributions from intensity mismatch and bandwidth smearing.}
    \begin{tabular}{c c c}
        \hline\hline
         & Warm & Cryogenic \\ \hline
        Null ratio $N$ & $1.4\pm2$\,\% & $2.4\pm3$\,\% \\
        Null stability $N_s$ & 0.037\,\%  & 0.057\,\% \\
        Intensity mismatch contribution $N_{\sigma_I}$  & $<0.05\,$\%  & $<0.05\,$\%  \\
        Bandwidth smearing contribution  & 0.135\,\%  & 0.135\,\%  \\ \hline
    \end{tabular}
    \label{tab:summary}
\end{table}
    
We observe a significant increase of the null ratio measured at cryogenic temperature compared to the experiment at ambient temperature. During the cryogenic experiment, more perturbations are visible from the output intensities in the left panel of Fig.\,\ref{fig:null_measure_cryo}, which are likely due to the nearby vacuum pump. The pump is placed on a different table, next to the optical bench, but its vibrations are propagating through the vacuum hose to the cryostat.
These perturbations are probably the reason for the higher null ratio and its higher standard deviation. Additional tests with the vacuum pump turned off should solve the issue.

\section{Conclusion}\label{sec:conclusion}
The first lab assembly of Asgard/NOTT is a benchmark for its integration within the Asgard instrument. Its main goals are threefold: (1) test and optimize the operation of all NOTT warm subsystems, (2) develop the software tools required to achieve deep interferometric nulls, (3) gain expertise and test the reliability of the assembling and alignment procedure.
Using the delay lines, and later a piezo-actuator, fringes were obtained between two of the injected beams and the null position with the deepest destructive interference could be reached. With a two beam recombination, a null of 1.4\,\% with a few $10^{-4}$ stability is obtained over 30\,s of measurements, which is consistent with previous studies done with a smaller bandpass \citep{SannyPhD}.
Finally, a test cryostat has been implemented on the bench to cool the chip down to 138\,K, and characterize its performance at cryogenic temperature. Cryogenic experiments show no significant evolution of the $\kappa$-matrix or of the splitting of the light inside the chip, and a larger null of 2.4\,\% is measured which is likely due to the vibrations induced by the connected vacuum pump. According to this experiment, the chip is compatible with cryogenic nulling operations for temperatures down to 138\,K and for the full L'-band passband. Future work will focus on implementing an anti-reflection coating on the input and output surfaces of the chip to increase its broadband throughput. Once installed, the final spectrograph will also enable spectral dispersion of the chip's outputs and spectral characterization of the beam-combiner's properties and splitting ratios at ambient and cryogenic temperatures down to 90\,K.

\subsection*{ACKNOWLEDGMENTS} 
SCIFY has received funding from the European Research Council (ERC); Award no. CoG –
866070 under the European Union’s Horizon 2020 research and innovation program.
R. L. has received funding from the Research Foundation - Flanders (FWO) under the grant number 1234224N. S. E., J. P. S., and T. A. S. acknowledge funding by the National Aeronautics and Space
Administration (NASA) through grant 80NSSC23K1473.

\bibliographystyle{unsrtnat}
\bibliography{references}  






\end{document}